\documentstyle[12pt]{article}
\begin{document}

\pagestyle{empty} \noindent
\hspace*{115mm} \normalsize hep-th/9301005 \\
 \hspace*{115mm} IASSNS-HEP-92/88 \\ \hspace*{115mm} December 1992 \\
\begin{center} \vspace*{30mm} \LARGE Duality, Marginal Perturbations and
Gauging
 \\
\vspace*{20mm} \large M\aa ns Henningson$^\star$ and Chiara R. Nappi$^\dagger$
 \\
\vspace*{10mm} \normalsize \it School of Natural Sciences \\ Institute for
 Advanced Study \\ Olden Lane \\ Princeton, NJ 08540 \\
\vspace*{30mm} \large Abstract \\ \end{center}
We study duality transformations for two-dimensional sigma models with abelian
 chiral isometries and prove that generic such transformations are equivalent
to
 integrated marginal perturbations by bilinears in the chiral currents, thus
 confirming a recent conjecture by Hassan and Sen formulated in the context of
 Wess-Zumino-Witten models. Specific duality transformations instead give rise
 to coset models plus free bosons.\\

\vspace*{10mm} \noindent
$\star$ Research supported by the Swedish Natural Science Research Council
 (NFR).\\
$\dagger$ Research supported in part by the Ambrose Monell Foundation.\\

\newpage \pagestyle{plain}
\begin{center} \large 1. Introduction \normalsize \\  \end{center}
Recently there has been a lot of interest in two-dimensional conformally
 invariant sigma models with abelian isometries. The space of theories with $d$
 abelian isometries transforms under a group of so called duality
 transformations, which is isomorphic to $O(d,d)$.These transformations are a
 generalization of the transformation introduced by Buscher \cite{Buscher} for
 the case of a single isometry. Buscher's transformation, in its turn, can be
 viewed as a generalization of the familiar $R \rightarrow 1/R$ symmetry in
 conformal field theory.

Duality transformations are intriguing and powerful symmetries, that may relate
  conformal string background with totally different spacetime geometries.
Indeed they have been used recently to generate new string solutions from
known ones. This paper is an attempt to understand in more detail some
 properties of this symmetry. In a recent
 paper \cite{Hassan-Sen} Hassan and Sen studied duality transformations of the
 Wess-Zumino-Witten model and related models. They found that the
 marginal perturbations by bilinears in the chiral currents of specific such
 models could be reproduced by suitable duality
 transformations, and they conjectured that this result should be generalizable
 to any Wess-Zumino-Witten model. In this paper we prove this conjecture in the
 more general context of sigma models with abelian chiral isometries. An
 important example is of course the Wess-Zumino-Witten model, since such a
 model based on the group $G$ possesses rank $G$ holomorphic and rank $G$
 anti-holomorphic abelian chiral isometries, but our considerations will be
more
 general. We will show that generic duality transformations indeed correspond
to
integrated marginal perturbations.

When this representation in terms of marginal deformations fails,
the duality transformation appear to be related to gauged models.
Transformations that relate a given model with chiral isometries to its gauged
 version
 plus a set of free bosons have already been discussed in the literature.
 Examples of such duality transformations
 have been given by
Kumar \cite{Kumar}, and by Ro\v cek and Verlinde \cite{Rocek-Verlinde} in the
 case of one holomorphic
 and one anti-holomorphic isometry.
 Here we investigate in detail this latter case, and find that specific duality
 transformations that yield models
 related  to the corresponding axial and vector coset models
are indeed the ones that cannot be represented in terms of marginal
perturbations by bilinears in the chiral currents. We conjecture that a
 similar result should be valid in general case. We hope that these
 investigations could be a step on the way to a better understanding of the
 moduli space of conformal field theories with not only abelian chiral
 isometries but more general chiral current algebras.

This paper is organized as follows: In section~2 we give a quick review of
sigma
 models with abelian isometries and the corresponding duality group. This
 discussion is specialized to models with chiral isometries in section~3. We
 show that a generic $O(d,d)$ transformation applied to such a model gives rise
 to a model of the same type with the same number of chiral
 isometries. In section~4 we show that infinitesimal duality transformations
 correspond to marginal perturbations. In section~5 we
 consider in full detail the simplest non-trivial example, i.~e. a model with
 one holomorphic and one
 anti-holomorphic isometry, and determine all models which are
 related to it by duality. Besides the previously mentioned models with chiral
 isometries we also find
 models which could be obtained by performing the coset construction on the
 original model plus a set of free bosons. This hints at a deeper relationship
 between duality and
 gauging, which we investigate in section~6 from a slightly different
 perspective.

\vspace*{15mm} \begin{center} \large 2. Abelian isometries and duality
 transformations \normalsize \\ \end{center}
In this section we give a brief review of sigma models with $d$ abelian
 isometries and the associated duality group $O(d,d)$. Readers are referred to
 the papers by Ro\v cek and Verlinde \cite{Rocek-Verlinde} and Giveon and Ro\v
 cek \cite{Giveon-Rocek} for more details.

We may choose coordinates so that the isometries act by translation of the
 coordinates $\theta = \left( \theta^1 \ldots \theta^d \right)$. The remaining
 coordinates are denoted $x^a$. The action may then be written in the form
\begin{equation}
S=\frac{1}{2\pi} \int d^2 z \; \left( \partial \theta E(x) \bar{\partial}
 \theta^t + \partial \theta F_{Ra}(x) \bar{\partial} x^a + \partial x^a
 F_{La}(x)  \bar{\partial} \theta^t \right) + S[x], \label{e1}
\end{equation}
where $E(x)$, $F_{Ra}(x)$ and $F_{La}(x)$ are matrices of type $d \times d$, $d
 \times 1$ and $1 \times d$ respectively. Matrix transposition is denoted $^t$.
 Henceforth we will often drop the $a$-index on $F_{Ra}(x)$ and $F_{La}(x)$.

The group $O(d,d)$ is defined as the set of all $2d \times 2d$ matrices $g$
that
 leave a metric $J_0$ of signature $d+d$ invariant:
\begin{equation}
g^t J_0 g = J_0,
\end{equation}
where
\begin{equation}
J_0 = \left( \begin{array}{cc} 0 & I \\ I & 0 \end{array} \right).
\end{equation}
Here $I$ is the $d$-dimensional identity matrix. If we decompose $g$ in block
 form as
\begin{equation}
g = \left( \begin{array}{cc} a & b \\ c & d \end{array} \right), \label{e2}
\end{equation}
where $a$, $b$, $c$ and $d$ are $d \times d$ matrices, this is equivalent to
demanding that
\begin{equation}
a^t c + c^t a = 0 \;\;\;\;\;\;\; b^t d + d^t b = 0 \;\;\;\;\;\;\; a^t d + c^t b
 = I.
\end{equation}

The $O(d,d)$ element $g$ in (\ref{e2}) acts on the sigma model defined by
 (\ref{e1}) by transforming it into a model of the same kind with $E(x)$,
 $F_R(x)$, $F_L(x)$ and $S[x]$ replaced by
\begin{eqnarray}
&& E^\prime(x) = \left( a E(x) + b \right) \left( c E(x) + d \right)^{-1}
 \nonumber\\
&& F_R^\prime(x) = \left( a - E^\prime(x)c \right) F_R(x) \label{e3}\\
&& F_L^\prime(x) = F_L(x) \left( c E(x) + d \right)^{-1} \nonumber\\
&& S^\prime[x] = S[x] - \frac{1}{2\pi} \int d^2 z \; \partial x^a F_{La}(x)
 \left( cE(x)+d \right)^{-1}cF_{Rb}(x) \bar{\partial} x^b \nonumber.
\end{eqnarray}
When accompanied by an appropriate shift of the dilaton field, as discussed in
 \cite{Buscher}, these transformations preserve conformal invariance of the
 model at one-loop level. Corrections are known to exist to all orders
 \cite{Sen} and preserve conformal invariance.

Matrices of the form
\begin{equation}
g = \left( \begin{array}{cc} (\alpha^t)^{-1} & 0 \\ 0 & \alpha \end{array}
 \right)
\end{equation}
constitute a $GL(d)$ subgroup of $O(d,d)$ that acts by linear coordinate
 transformations among the $\theta^i$ coordinates. The matrices
\begin{equation}
g = \left( \begin{array}{cc} I & \beta \\ 0 & I \end{array} \right) \;\;\;\;\;
 \beta + \beta^t = 0
\end{equation}
form a ${\bf R}^{d(d-1)/2}$ subgroup which corresponds to adding total
 derivative terms to the action. Together these elements generate a subgroup
 $\Lambda(d)$ of elements of the form
\begin{equation}
\lambda = \left( \begin{array}{cc} (\alpha^t)^{-1} & \beta \\ 0 & \alpha
 \end{array} \right) \;\;\;\;\; \alpha^t \beta + \beta^t \alpha = 0, \label{e5}
\end{equation}
which act trivially on (\ref{e1}) in the sense that the transformed model is
 equivalent to the original one up to coordinate transformations and partial
 integrations.

\vspace*{15mm} \begin{center} \large 3. Sigma models with chiral isometries
 \normalsize \\ \end{center}
We now specialize the discussion to the case where the $d$ isometries may be
 decomposed as $d_L$ holomorphic and $d_R$ anti-holomorphic chiral isometries
 with $d=d_L+d_R$. This will allow us to make more specific statements about
the
 dual models. We will see that for a generic $O(d,d)$ transformation the dual
 model possesses the same number of holomorphic and anti-holomorphic chiral
 isometries.

With an appropriate choice of coordinates, the action may be written in the
form
 (\ref{e1}) with
\begin{equation}
\theta = \left( \begin{array}{cc} \theta_R & \theta_L \end{array} \right) =
 \left( \theta_R^1 \ldots \theta_R^{d_R} \;\; \theta_L^1 \ldots \theta_L^{d_L}
 \right)
\end{equation}
and
\begin{equation}
E(x) = \left( \begin{array}{cc} I_R & 2B(x) \\ 0 & I_L \end{array} \right)
 \;\;\;\;\; F_R(x) = \left( \begin{array}{c} G_R(x) \\ 0 \end{array} \right)
 \;\;\;\;\;  F_L(x) = \left( \begin{array}{cc} 0 & G_L(x) \end{array} \right).
 \label{e10}
\end{equation}
Here $I_L$ and $I_R$ denote the $d_L$- and $d_R$-dimensional identity matrices,
 and $B(x)$, $G_R(x)$ and $G_L(x)$ are matrices of type $d_R \times d_L$, $d_R
 \times 1$ and $1 \times d_L$ respectively.

Written out explicitly, this action is
\begin{eqnarray}
S_{LR} & = & \frac{1}{2\pi} \int d^2 z \; \left( \partial \theta_L
 \bar{\partial} \theta_L^t + \partial \theta_R \bar{\partial} \theta_R^t +
 \partial \theta_R 2 B(x) \bar{\partial} \theta_L^t \right. \label{e4}\\
&& + \left. \partial \theta_R G_{Ra}(x) \bar{\partial} x^a + \partial x^a
 G_{La}(x) \bar{\partial} \theta_L \right) + S[x]. \nonumber
\end{eqnarray}
The equations of motion that follow from a variation of $\theta_R$ and
 $\theta_L$ are $\partial \bar{J}_R = 0$ and $\bar{\partial} J_L = 0$
 respectively, where the chiral currents are given by
\begin{eqnarray}
\bar{J}_R & = & \left( \bar{J}^1_R \ldots \bar{J}^{d_R}_R \right) =
 \bar{\partial} \theta_R + \bar{\partial} \theta_L B(x)^t + \frac{1}{2}
 G_{Ra}(x)^t \bar{\partial} x^a \label{e20}\\
J_L & = & \left( J^1_L \ldots J^{d_L}_L \right) = \partial \theta_L + \partial
 \theta_R B(x) + \frac{1}{2} \partial x^a G_{La}(x). \nonumber
\end{eqnarray}
The conformal dimension of the holomorphic (anti-holomorphic) current $J_L$
 ($\bar{J}_R$) is $(1,0)$ ($(0,1)$).

Our object is to analyze the orbit of the action (\ref{e4}) under $O(d,d)$
 acting as in (\ref{e3}). However, we are only interested in classically
 inequivalent models, which could not be obtained from one another by
coordinate
 transformations and partial integrations, so we should rather consider the
 right coset $\Lambda(d) \backslash O(d,d)$, where the subgroup $\Lambda(d)$
was
 defined in (\ref{e5}). Furthermore, there is a subgroup $\Omega(d_L,d_R)$ of
 $O(d,d)$ elements that leave the action (\ref{e4}) invariant. We will
construct
 this subgroup explicitly in a simple case in section~5. Our real object of
 interest is therefore the double coset $\Lambda(d) \backslash O(d,d) /
 \Omega(d_L,d_R)$, i.~e. the set of equivalence classes of $O(d,d)$ under the
 equivalence relation
\begin{equation}
g \sim \lambda g \omega \;\; ; \;\;\; \lambda \in \Lambda(d), \;\; \omega \in
 \Omega(d_L,d_R).
\end{equation}

To analyze this coset, let us first consider $O(d,d)$ elements such that the
 submatrix $d$ in (\ref{e2}) is invertible, i.~e. $\det d \neq 0$. This is
 certainly true in a neighbourhood of the identity of $O(d,d)$, and for the
 simplest non-trivial case $d_L=d_R=1$ we have checked that all $O(2,2)$
 elements of the $\det d = 0$ type are equivalent to elements with $\det d \neq
 0$  modulo $\Omega(1,1)$ acting from the right. We conjecture that also in the
 general case one needs only to consider elements of $O(d,d)$ with $\det d \neq
 0$, although we have no proof of this.

We may parametrize an $O(d,d)$ element of the form (\ref{e2}) with $\det d \neq
 0$ by the $d \times d$ matrices $d$, $e$ and $f$, where $e$ and $f$ are
defined
 by
\begin{equation}
e = d^{-1}c \;\;\;\;\;\;\;\; f = b^t d
\end{equation}
so that
\begin{eqnarray}
g = \left( \begin{array}{cc} a & b \\ c & d \end{array} \right) = \left(
 \begin{array}{cc} (d^t)^{-1} (I-fe) & (d^t)^{-1} f^t \\ de & d \end{array}
 \right). \label{e6}
\end{eqnarray}
The requirement that $g$ be an element of $O(d,d)$ amounts to $e$ and $f$ being
 antisymmetric;
\begin{equation}
e+e^t=f+f^t=0,
\end{equation}
while $d$ is unconstrained, apart from the requirement of invertibility.

Multiplication of (\ref{e6}) from the left by an element $\lambda \in
 \Lambda(d)$ of the form (\ref{e5}) yields a new element $g^\prime = \lambda g$
 of the form (\ref{e6}) with $d$, $e$ and $f$ replaced by
\begin{equation}
d^\prime = \alpha d \;\;\;\;\;\;\; e^\prime = e \;\;\;\;\;\;\; f^\prime = f +
 d^t \beta^t \alpha d.
\end{equation}
We see that $e$ is invariant under such transformations, and furthermore we may
 transform $d$ and $f$ to any preferred values $d_0$ and $f_0$ (fulfilling
$\det
 d_0 \neq 0$ and $f_0 + f_0^t = 0$) by choosing
\begin{equation}
\alpha = d_0 d^{-1} \;\;\;\;\;\;\; \beta = (d_0^t)^{-1} (f_0^t-f^t) d^{-1}.
\end{equation}
The equivalence classes of $O(d,d)$ modulo $\Lambda (d)$ may thus be labeled by
 the matrix $e$, which is only subject to the constraint of being
antisymmetric.

In the case at hand, a convenient choice of representative in (almost) every
 equivalence class may be described as follows. Introduce the $d \times d$
 matrix $J$ as
\begin{equation}
J = \left( \begin{array}{cc} I_R & 0 \\ 0 & - I_L \end{array} \right),
\end{equation}
and define $h$ by
\begin{equation}
h = (J-e)(J+e)^{-1}, \label{e13}
\end{equation}
which may be inverted to yield
\begin{equation}
e = (I+h)^{-1} (I-h) J. \label{e13b}
\end{equation}
These relations are well defined for generic $e$ and $h$ and at least in a
 neighbourhood of $h=I$, $e=0$. We will come back to the remaining cases where
 $J+e$ is not invertible in section~5. It is easy to show that the antisymmetry
 of $e$ is equivalent to $h$ being an element of the group $O(d_L,d_R)$, i.~e.
\begin{equation}
h^t J h = J. \label{e13c}
\end{equation}
We now choose
\begin{equation}
d = \frac{1}{2} (I+h) \;\;\;\;\;\;\;\; f = \frac{1}{4} (I-h^t) J (I+h).
 \label{e7}
\end{equation}
The property (\ref{e13c}) implies that $f$ is antisymmetric as required.

To see the advantage of this choice we write $h$ in block form
\begin{equation}
h = \left( \begin{array}{cc} V_R & T \\ S & V_L \end{array} \right),
\label{e14}
\end{equation}
where $V_R$, $V_L$, $S$ and $T$ are matrices of type $d_R \times d_R$, $d_L
 \times d_L$, $d_L \times d_R$ and $d_R \times d_L$ respectively. The
 requirement (\ref{e13c}) that $h$ be an element of $O(d_L,d_R)$ amounts to
\begin{equation}
V_R^t V_R - S^t S = I_R \;\;\;\;\;\;\; V_L^t V_L - T^t T = I_L \;\;\;\;\;\;\;
 V_R^t T - S^t V_L = 0. \label{e11}
\end{equation}
Inserting this $h$ in (\ref{e13b}), (\ref{e7}) and (\ref{e6}) we get
\begin{equation}
\begin{array}{ll} a = \frac{1}{2} \left( \begin{array}{cc} I_R + V_R & - T \\ -
 S & I_L + V_L \end{array} \right) &  b = \frac{1}{2} \left( \begin{array}{cc}
 I_R - V_R & - T \\ S & - I_L + V_L \end{array} \right) \\ & \\
c = \frac{1}{2} \left( \begin{array}{cc} I_R - V_R & T \\ - S & - I_L + V_L
 \end{array} \right) & d = \frac{1}{2} \left( \begin{array}{cc} I_R + V_R & T
\\
 S & I_L+ V_L \end{array} \right) \end{array}. \label{e13d}
\end{equation}

We now apply the transformations (\ref{e3}), with $a$, $b$, $c$ and $d$ given
by
 (\ref{e13d}), to the model defined by (\ref{e1}) and (\ref{e10}) and get
\begin{eqnarray}
&& E^\prime (x) = \left( \begin{array}{cc} I_R & 2 B^\prime (x) \\ 0 & I_L
 \end{array} \right) \;\;\;\; F_R^\prime (x) = \left( \begin{array}{c}
 G_R^\prime (x) \\ 0 \end{array} \right) \;\;\;\; F_L^\prime (x) = \left(
 \begin{array}{cc} 0 & G_L^\prime (x) \end{array} \right) \nonumber\\
&& S^\prime [x] = S [x] + \frac{1}{2\pi} \int d^2 z \; \frac{1}{2} \partial x^a
 G_{La}(x) (V_L - S B(x) )^{-1} S G_{Rb}(x) \bar{\partial} x^b,
\end{eqnarray}
where
\begin{eqnarray}
&& B^\prime (x)  = (V_R B(x) - T) (V_L - S B(x))^{-1} \nonumber\\
&& G_R^\prime (x) = (V_R^t - B(x) T^t)^{-1} G_R(x) \label{e12}\\
&& G_L^\prime (x) = G_L (x) (V_L - S B(x))^{-1}. \nonumber
\end{eqnarray}
We see that the matrices $E^\prime(x)$, $F_R^\prime(x)$ and $F_L^\prime(x)$ are
 still of the form (\ref{e10}), and the transformed model is thus of the same
 kind as the original one, with $d_L$ holomorphic and $d_R$ anti-holomorphic
 chiral isometries. The only restrictions that we have imposed on the $O(d,d)$
 transformation to reach this result is that the submatrix $d$ in (\ref{e2})
and
 the matrix $J+e$ in (\ref{e13}) are invertible.

\vspace*{15mm} \begin{center} \large 4. Infinitesimal transformations and
 marginal perturbations \normalsize \\ \end{center}
To get a better understanding of the transformations in the previous section,
we
 will here consider infinitesimal transformations $h = I + \epsilon \tilde{h} +
 {\cal O} ( \epsilon^2)$ with
\begin{equation}
\tilde{h} = \left( \begin{array}{cc} \tilde{V}_R & \tilde{T} \\ \tilde{S} &
 \tilde{V}_L \end{array} \right).
\end{equation}
The constraints (\ref{e11}) give
\begin{equation}
\tilde{V}_R + \tilde{V}_R^t = 0 \;\;\;\;\;\;\; \tilde{V}_L + \tilde{V}_L^t = 0
 \;\;\;\;\;\;\; \tilde{T} - \tilde{S}^t = 0.
\end{equation}
Inserting this $h$ in (\ref{e12}) yields the infinitesimal transformation
\begin{eqnarray}
&& B^\prime (x) = B(x) + \epsilon \left( \tilde{V}_R B(x) - B(x) \tilde{V}_L -
 \tilde{S}^t + B(x) \tilde{S} B(x) \right) + {\cal O}(\epsilon^2) \nonumber\\
&& G_R^\prime (x) = G_R(x) + \epsilon \left( - \tilde{V}_R + B(x) \tilde{S}
 \right) G_R (x) + {\cal O}(\epsilon^2) \label{e15}\\
&& G_L^\prime (x) = G_L(x) + \epsilon G_L(x) \left( - \tilde{V}_L + \tilde{S}
 B(x) \right) + {\cal O}(\epsilon^2) \nonumber\\
&& S^\prime [x] = S[x] + \frac{\epsilon}{2} \frac{1}{2\pi} \int d^2 z \;
 \partial x^a G_{La}(x) \tilde{S} G_{Rb}(x) \bar{\partial} x^b + {\cal
 O}(\epsilon^2). \nonumber
\end{eqnarray}

These transformations can be interpreted as a marginal perturbation by a
 bilinear in the holomorphic and anti-holomorphic chiral currents (\ref{e20}),
 as we will now show. For an arbitrary $d_L \times d_R$ matrix $\tilde{S}$, the
 operator $J_L \tilde{S} \bar{J}_R^t$ is (classically) of dimension $(1,1)$ and
 may be added as a perturbation to the Lagrangian (\ref{e4}). One finds that
\begin{eqnarray}
S^\prime_{LR} & = & S_{LR} + 2 \epsilon \frac{1}{2\pi} \int d^2 z \; J_L
 \tilde{S} \bar{J}_R^t = \nonumber\\
& = & \frac{1}{2\pi} \int d^2 z \; \left( \partial \theta_R ( I_R + 2 \epsilon
 B(x) \tilde{S} ) \bar{\partial} \theta_R^t + \partial \theta_L ( I_L+ 2
 \epsilon \tilde{S} B(x) ) \bar{\partial} \theta_L^t \right. \label{e15b}\\
&& + \left. \partial \theta_R ( 2 B(x) + 2 \epsilon B(x) \tilde{S} B(x) )
 \bar{\partial} \theta_L^t + \partial \theta_L 2 \epsilon \tilde{S}
 \bar{\partial} \theta_R^t + \right. \nonumber\\
& & + \left. \partial \theta_R ( I_R + \epsilon B(x) \tilde{S} ) G_{Ra}(x)
 \bar{\partial} x^a + \partial x^a G_{La}(x) ( I_L + \epsilon \tilde{S} B(x) )
 \bar{\partial} \theta_L^t \right. \nonumber\\
& & + \left. \partial \theta_L \epsilon \tilde{S} G_{Ra}(x) \bar{\partial} x^a
+
 \partial x^a G_{La}(x) \epsilon \tilde{S} \bar{\partial} \theta_R^t +
 \frac{\epsilon}{2} \partial x^a G_{La}(x) \tilde{S} G_{Rb}(x) \bar{\partial}
 x^b \right) + S[x]. \nonumber
\end{eqnarray}
After an infinitesimal coordinate change
\begin{eqnarray}
\theta_R & \rightarrow & \theta_R + \epsilon \theta_R \tilde{V}_R - \epsilon
 \theta_L \tilde{S} \\
\theta_L & \rightarrow & \theta_L + \epsilon \theta_L \tilde{V}_L - \epsilon
 \theta_R \tilde{S}^t \nonumber
\end{eqnarray}
with $\tilde{V}_R+\tilde{V}_R^t=0$ and $\tilde{V}_L+\tilde{V}_L^t=0$ we get
 $S_{LR}^\prime$ of the form (\ref{e4}) with $B^\prime (x)$, $G_R^\prime(x)$,
 $G_L^\prime(x)$ and $S^\prime [x]$ given by (\ref{e15}). The infinitesimal
 duality transformations (\ref{e15}) are thus equivalent to marginal
 perturbations (\ref{e15b}). Since the marginally perturbed model has the same
 number of abelian chiral isometries, we can repeat the process. The result of
 applying such an "integrated" marginal perturbation to the model (\ref{e4}) is
 a model of the form (\ref{e12}), which is obtained by a finite $O(d,d)$
 transformation.

Note that this relationship between duality transformations and marginal
 perturbations provides a simple check that the former preserve conformal
 invariance. Indeed, Chaudhuri and Schwarz \cite{Chaudhuri-Schwarz} have proved
 that marginal perturbations by a bilinear in commuting chiral currents
preserve conformal invariance.

\vspace*{15mm} \begin{center} \large 5. The case of $d_L=d_R=1$ \normalsize \\
 \end{center}
In section~3 we have seen that generic duality transformations of the model
 (\ref{e4}) yield a model of the same type with the couplings given in
 (\ref{e12}). In this section we will examine the simplest non-trivial example
 with one holomorphic and one anti-holomorphic isometry in somewhat more detail
 to determine the complete orbit of (\ref{e4}) under duality transformations.

We begin our investigations by determining the group $\Omega(1,1)$ of $O(2,2)$
 elements that leave the action $S_{LR}$ invariant under the transformations
 (\ref{e3}). This turns out to be an abelian discrete group with four elements:
\begin{equation}
\begin{array}{cc} \omega_1 = \left( \begin{array}{cccc} 1 & 0 & 0 & 0 \\ 0 & 1
&
 0 & 0 \\ 0 & 0 & 1 & 0 \\ 0 & 0 & 0 & 1 \end{array} \right) & \omega_2 =
\left(
 \begin{array}{cccc} 0 & 0 & -1 & 0 \\ 0 & 0 & 0 & 1 \\ -1 & 0 & 0 & 0 \\ 0 & 1
 & 0 & 0 \end{array} \right) \\ & \\
\omega_3 = \left( \begin{array}{cccc} 1 & 0 & 0 & 0 \\ 0 & 0 & 0 & 1 \\ 0 & 0 &
 1 & 0 \\ 0 & 1 & 0 & 0 \end{array} \right) & \omega_4 = \left(
 \begin{array}{cccc} 0 & 0 & -1 & 0 \\ 0 & 1 & 0 & 0 \\ -1 & 0 & 0 & 0 \\ 0 & 0
 & 0 & 1 \end{array} \right) \end{array}. \label{e50}
\end{equation}

We have already mentioned that all elements $g$ of $O(2,2)$ with the
determinant
 of the submatrix $d$ in (\ref{e2}) vanishing are equivalent modulo
 $\Omega(1,1)$ acting from the right to elements with $\det d \neq 0$. We need
 therefore only consider the case $\det d \neq 0$. From our previous reasoning
 we know that the equivalence classes of such $O(2,2)$ elements modulo $\Lambda
 (2)$ acting from the left may be labeled by the $2 \times 2$ antisymmetric
 matrix $e$ given by
\begin{equation}
e = d^{-1} c = \left( \begin{array}{cc} 0 & -x \\ x & 0 \end{array} \right).
\end{equation}
Here $x$ may take any real value, but the transformation $g \rightarrow g
 \omega_2$, with $\omega_2$ given in (\ref{e50}), induces
\begin{equation}
e \rightarrow \left( \begin{array}{cc} 0 & -x^{-1} \\ x^{-1} & 0 \end{array}
 \right),
\end{equation}
so we may restrict our attention to the interval $-1 \leq x \leq 1$.

For $-1<x<1$ we may use (\ref{e13}) to get
\begin{equation}
h = \left( \begin{array}{cc} \cosh t & - \sinh t \\ - \sinh t & \cosh t
 \end{array} \right),
\end{equation}
where $x$ and $t$ are related by $x = (1 + \cosh t )^{-1} \sinh t$ and $-\infty
 < t < \infty$. The corresponding transformations act on the model (\ref{e4})
 via (\ref{e14}) and (\ref{e12}) as
\begin{eqnarray}
&& B^\prime (x) = (\cosh t + B(x) \sinh t)^{-1} (\sinh t + B(x) \cosh t)
 \nonumber\\
&& G_R^\prime (x) = (\cosh t + B(x) \sinh t)^{-1} G_R (x) \label{e41}\\
&& G_L^\prime (x) = (\cosh t + B(x) \sinh t)^{-1} G_L (x) \nonumber\\
&& S^\prime [x] = S[x] - \frac{1}{2\pi} \int d^2 z \; \frac{1}{2} (\cosh t +
 B(x) \sinh t)^{-1} \sinh t \partial x^a G_{La}(x) G_{Rb}(x) \bar{\partial}
x^b.
 \nonumber
\end{eqnarray}
As before, this transformation preserves the number of chiral isometries
 $d_L=d_R=1$.

The two remaining values $x = \pm 1$ mean that $J+e$ is not invertible so that
 (\ref{e13}) may not be used. Instead we can choose the representatives
\begin{equation}
\begin{array}{cc} a = \frac{1}{2} \left( \begin{array}{cc} -1 & \mp 1 \\ -1 &
 \pm 1 \end{array} \right) & b = \frac{1}{2} \left( \begin{array}{cc} -1 & \pm
1
 \\ 1 & \pm 1 \end{array} \right) \\ & \\
c = \frac{1}{2} \left( \begin{array}{cc} -1 & \pm 1 \\ 1 & \pm 1 \end{array}
 \right) & d = \frac{1}{2} \left( \begin{array}{cc} -1 & \mp 1 \\ -1 & \pm 1
 \end{array} \right) \end{array},
\end{equation}
which, when inserted in (\ref{e3}) together with (\ref{e10}), yield
\begin{eqnarray}
&& E^\prime (x) = \left( \begin{array}{cc} 1 & 0 \\ 0 & \frac{1 \mp B(x)}{1 \pm
 B(x)} \end{array} \right) \;\;\;\;\; F_R^\prime(x) = \left( \begin{array}{c} 0
 \\ - \frac{G_R(x)}{1 \pm B(x)} \end{array} \right) \;\;\;\;\; F_L^\prime(x) =
 \left( \begin{array}{cc} 0 & \frac{G_L(x)}{B(x) \pm 1} \end{array} \right)
 \nonumber\\
&& S^\prime [x] = S[x] - \frac{1}{2\pi} \int d^2 z \; \frac{1}{2} \frac{1}{B(x)
 \pm 1} \partial x^a G_{La}(x) G_{Rb}(x) \bar{\partial} x^b.
\end{eqnarray}
These theories thus consist of a free boson plus a sigma model which is in fact
 the coset model $S_V$ or $S_A$ described by Ro\v cek and Verlinde
 \cite{Rocek-Verlinde}:
\begin{eqnarray}
S_{V/A} & = & \frac{1}{2\pi} \int d^2 z \; \left( \frac{1 \mp B(x)}{1 \pm B(x)}
 \partial \theta \bar{\partial} \theta - \frac{G_{Ra}(x)}{1 \pm B(x)} \partial
 \theta \bar{\partial} x^a \pm \frac{G_{La}(x)}{1 \pm B(x)} \partial x^a
 \bar{\partial} \theta \right. \nonumber\\
&& \mp \left. \frac{1}{2} \frac{G_{La}(x) G_{Rb}(x)}{1 \pm B(x)} \partial x^a
 \bar{\partial} x^b \right) + S[x]. \label{e33}
\end{eqnarray}
It is probably true that all $O(d,d)$ transformations such that $J+e$ is not
 invertible give rise to theories with free bosons. In fact for the case where
 $d_L=d_R=d/2$, Kumar \cite{Kumar} has given an explicit $O(d,d)$
 transformation, which is such that $J+e$ is not invertible, and which
 transforms the model (\ref{e4}) into a model with $d/2$ free bosons.

Note that these models in a sense also have $d_L$ holomorphic and $d_R$
 anti-holomorphic abelian isometries, although in this case the holomorphic
 isometries are really pairwise identical to the anti-holomorphic ones, both
 acting as translations of the $d/2$ free bosons. We conjecture that, with this
 definition of the number of chiral isometries, all models obtained by $O(d,d)$
 transformations of (\ref{e4}) will have $d_L$ holomorphic and $d_R$
 anti-holomorphic abelian isometries.

\vspace*{15mm} \begin{center} \large 6. Gauging and duality \normalsize \\
 \end{center}
In this section we will discuss more explicitly the relation
 between quotients, quotients by chiral currents and duality in the $d_L=d_R=1$
 case. First of all, to clarify the constructions in the previous
 section, we should note that all we have done there is
 dualizing a combination of holomorphic and anti-holomorphic isometries labeled
 by a mixing angle $\alpha$. A way to see this is to change variables
 in  (\ref{e4}) from $\theta_L$ and $\theta_R$ to $\theta_0$ and
 $\theta_1$ defined as
\begin{equation}
\left( \begin{array}{c} \theta_0 \\ \theta_1 \end{array} \right) = \left(
 \begin{array}{cc} \cos \alpha  & \sin \alpha \\ - \sin \alpha & \cos \alpha
 \end{array} \right) \left( \begin{array}{c} \theta_L \\ \theta_R \end{array}
 \right).
\end{equation}
We then perform a duality transformation with respect to the isometry which
acts
 as a translation of the $\theta_0$ coordinate, according to the
 prescription of Buscher \cite{Buscher}.
Namely, one goes to a first order form for (\ref{e4}) by introducing $V_\mu=
\partial_\mu\theta_0$ via a lagrange multiplier, and then solves
 for $V_\mu$ through its field equations.
 The explicit action for the dual model is then
\begin{eqnarray}
S_{dual} & = & \frac{1}{2\pi} \int d^2 z \; \frac{1}{1 + B(x) \sin 2 \alpha}
 \left( \partial \theta_0 \bar{\partial} \theta_0 + \partial \theta_1
 \bar{\partial} \theta_1  + (1 + \cos 2 \alpha ) B(x) \partial \theta_1
 \bar{\partial} \theta_0 \right. \nonumber\\
&& + (1-\cos 2 \alpha) B(x) \partial \theta_0 \bar{\partial} \theta_1 +
 G_{La}(x) \partial x^a ( \cos \alpha \bar{\partial} \theta_0 - \sin
 \alpha \bar{\partial} \theta_1) \label{e37}\\
&& \left. + G_{Ra}(x) \bar{\partial} x^a ( \cos \alpha
 \partial \theta_1 - \sin \alpha \partial \theta_0 ) \right) + S^\prime [x].
 \nonumber
\end{eqnarray}
with
\begin{equation}
S'[x] = S[x] - \frac{1}{2} \int d^2z \; {{\sin 2\alpha}
\over{1+B(x)\sin 2\alpha}}\partial x^a G_{La}(x)G_{Rb}(x){\bar\partial
x^b}. \label{e107}
\end{equation}

To turn (\ref{e37}) into a model of the left-right symmetric type as in
(\ref{e4}), we need to make
the further change of variables
\begin{equation}
\left( \begin{array}{c} \theta_0 \\ \theta_1 \end{array} \right) = \left(
 \begin{array}{cc} \sin \alpha & \cos \alpha \\ \cos \alpha & \sin \alpha
 \end{array} \right) \left( \begin{array}{c} \theta_R^\prime \\ \theta_L^\prime
 \end{array} \right),
\end{equation}
which is non-singular for $\cos 2 \alpha \neq 0$, i.~e. for $- \pi/4 < \alpha <
 \pi/4$. Then (\ref{e37}) yields the chiral model (\ref{e41}) with
 $\tanh t = \sin 2 \alpha$. The
 values $\alpha = 0$ and $\alpha = \pi /2$, corresponding to dualizing a chiral
 isometry, leave the
 model invariant, as was noted in \cite{Rocek-Verlinde}. The corresponding
 $O(2,2)$ transformations, their product and the identity constitute the
 subgroup $\Omega(1,1)$ discussed in the previous section. For the values of
 $\alpha = \pm \pi/4$,  instead, the change of
 variables to ${\theta_0}'$ and ${\theta_1}'$ defined by
\begin{equation}
\left( \begin{array}{c} \theta_0 \\ \theta_1 \end{array} \right) =
\frac{1}{\sqrt{2}} \left( \begin{array}{cc} \pm 1 & \pm 1 \\ 1 & -1 \end{array}
\right) \left( \begin{array}{c} {\theta_0}' \\ {\theta_1}' \end{array} \right)
\end{equation}
leads to free boson ${\theta_0}'$ plus $S_V$ or $S_A$ coset theories
(\ref{e33})
   , as
 already discussed in the previous section, and noticed originally in
\cite{Rocek-Verlinde}.

 Before rederiving (\ref{e37}) in a slightly different, although probably more
enlightening way,
 we will discuss gauging of sigma models with left and right chiral isometries.
 Starting from the action (\ref{e4}) we can
 gauge any combination of left and right isometries, parametrized by a mixing
 angle $\alpha$, by using the minimal coupling prescription
\begin{equation}
\partial \theta_R \rightarrow \partial \theta_R + \frac{1}{\sqrt{2}} A \cos
 \alpha \;\;\;\;\;\;\; \partial \theta_L \rightarrow \partial \theta_L +
 \frac{1}{\sqrt{2}} A \sin \alpha \label{e100}
\end{equation}
and analogously for the anti-holomorphic partial derivatives. This gives the
 gauged action
\begin{equation}
S_{gauged} = S_{LR} + \frac{1}{2\pi} \int d^2 z \; \left( \frac{1}{2} A \bar{A}
   ( 1+
 B(x) \sin 2\alpha ) + \frac{1}{\sqrt{2}} A \bar{L}_R + \frac{1}{\sqrt{2}} L_L
 \bar{A} \right), \label{e30}
\end{equation}
where
\begin{eqnarray}
&& \bar{L}_R = \sin \alpha \bar{\partial} \theta_L + \cos \alpha \bar{\partial}
 \theta_R + 2 B(x) \cos \alpha \bar{\partial} \theta_L + \cos \alpha G_{Ra}(x)
 \bar{\partial} x^a \label{e200}\\
&& L_L = \sin \alpha \partial \theta_L + \cos \alpha \partial \theta_R + 2 B(x)
 \sin \alpha \partial \theta_R + \sin \alpha G_{La}(x) \partial x^a. \nonumber
\end{eqnarray}
After integrating out the gauge field we get the action
\begin{eqnarray}
S_{gauged} & = & S_{LR} - \frac{1}{2\pi} \int d^2 z \; \frac{L_L \bar{L}_R}{1
 + B(x) \sin 2 \alpha} \nonumber\\
& = & \frac{1}{2\pi} \int d^2 z \; \left( 1 + B(x) \sin 2 \alpha \right)^{-1}
 \label{e32}\\
&& \left( \partial \theta \bar{\partial} \theta - \sin \alpha \partial \theta
 G_{Ra}(x) \bar{\partial} x^a + \cos \alpha G_{La} \partial x^a \bar{\partial}
 \theta \right) + S^\prime [x], \nonumber
\end{eqnarray}
where $S^\prime [x]$ is given in (\ref{e107}) and $\theta = \cos \alpha
\theta_L
 - \sin \alpha \theta_R$ is the gauge invariant linear combination of
$\theta_L$
 and $\theta_R$.
The above gauging procedure is fully gauge invariant for any $\alpha$, i.~e.
for
 any combination of left and right isometries. However, it is not a very
interesting one, since it does not automatically lead to conformal theories,
 even if  the starting theory is conformal. From this point of view,
 a slightly different procedure,  that mimics the coset construction and
  allows  to couple the gauge field directly to the chiral currents,
seems more interesting.
It is possible only for
 specific "anomaly-free" combinations of holomorphic and anti-holomorphic
 isometries. If the starting model is conformally invariant, the coset model
 obtained by this procedure is  probably conformally invariant as well,
 as it certainly is in the  case of the gauged Wess-Zumino-Witten model.

This new procedure amounts to the addition of the term
\begin{equation}
\frac{1}{2\pi} \int d^2 z \; \frac{1}{\sqrt{2}} ( \cos \alpha \theta_R - \sin
 \alpha \theta_L ) ( \partial \bar{A} - \bar{\partial} A ) \label{e31}
\end{equation}
to (\ref{e30}), which completes $L_L$ and $\bar{L}_R$ into the chiral currents
 $J_L$ and $\bar{J}_R$ defined in (\ref{e20}). The resulting gauged action is
\begin{equation}
S_{coset} = S_{LR} + \frac{1}{2\pi} \int d^2 z \; \left( \frac{1}{2} A \bar{A}
 ( 1 + B(x) \sin 2 \alpha ) + \sqrt{2} \cos \alpha A \bar{J}_R + \sqrt{2} \sin
 \alpha \bar{A} J_L \right) . \label{e300}
\end{equation}
Upon integrating out the gauge field this yields
\begin{equation}
S_{coset} = S_{LR} - \frac{1}{2\pi} \int d^2 z \; \frac{2 \sin 2 \alpha}{1 +
 B(x) \sin 2 \alpha} J_L \bar{J}_R. \label{e35}
\end{equation}
However, this procedure is legitimate only for $\alpha = \pm \pi /4$,  since
the
 term (\ref{e31}) is gauge invariant only for these values. In this case we get
\begin{equation}
S_{V/A} = S_{LR} \mp \frac{1}{2\pi} \int d^2 z \; \frac{2}{1 \pm B(x)} J_L
 \bar{J}_R.
\end{equation}
These theories are exactly the axial and vector quotients (\ref{e33}).
  Notice that gauging by chiral currents (\ref{e35}) and plain
 gauging (\ref{e32})  yield different results.
 Indeed, at $\alpha = \pm \pi /4$ (\ref{e32})
 and (\ref{e33}) are different. In both cases, gauge invariance of course leads
 to a reduction of degrees of freedom, and only the gauge invariant combination
 $\theta = \theta_R - \theta_L$ survives.

Getting back to the dual action ({\ref{e37}), an alternative way to obtain it
is
 by gauging any combination of left and right isometries in (\ref{e4})
 by minimal coupling as in (\ref{e100}) and  introducing a Lagrange multiplier
of the form  $\frac{1}{2\pi} \phi( {\bar\partial}A - \partial{\bar A} )$.
The gauged action is then
\begin{equation}
S_{dual} = S_{LR} + \frac{1}{2\pi} \int d^2 z \; \left( \frac{1}{2} A
 \bar{A} ( 1 \! + \! B(x) \sin 2\alpha ) + \frac{1}{\sqrt{2}} A (\bar{L}_R -
 {\bar\partial}\phi)+ \frac{1}{\sqrt{2}} (L_L + \partial\phi)  \bar{A} \right).
 \label{e110}
\end{equation}
 After integrating out the  gauge fields, one gets  the action
\begin{equation}
S_{dual} = S_{LR} - {1\over {2\pi}}\int d^2z
{1\over{1+B\sin 2\alpha}}({\bar L}_R - {\bar\partial}
\phi) (L_L + \partial\phi) \label{e111}
\end{equation}
The part of (\ref{e111}) independent of $\phi$ is  exactly the gauged
 lagrangian (\ref{e32}), and one recognizes that it is the
 part of the dual lagrangian (\ref{e37})  that depends on $\theta_0 =\theta$
only. One can furthermore
 show that the terms in (\ref{e111}) that depend on the lagrange
 multiplier $\phi$ give rise to the remaining pieces in (\ref{e37}), with
the identification $\phi=\theta_1$.
So dualizing can be interpreted as gauging in the presence of a lagrange
multiplier.

Obviously, one can add to (\ref{e110})  any term of the form (\ref{e31}),
 whether it is gauge invariant
 or not, since it can always be absorbed in a shift of the lagrange multiplier
\begin{equation}
\phi \rightarrow \phi' = \phi +
 \frac{1}{\sqrt{2}} ( \cos \alpha \theta_R - \sin
 \alpha \theta_L ) \label{e103}
\end{equation}
The lagrangian obtained by (\ref{e110}) after the addition of (\ref{e31})
 can always be made gauge invariant, by choosing $\phi$ to transform
 properly under gauge transformation in such a way to compensate the change of
(\ref{e31}). The answer will still be (\ref{e37}).

Moreover, for $\alpha \neq \pm{\pi\over 4}$ we can also fix the gauge by
setting
    $\phi = 0$, or
\begin{equation}
\phi' =
 \frac{1}{\sqrt{2}} ( \cos \alpha \theta_R - \sin
 \alpha \theta_L ). \label{e104}
\end{equation}
Then the gauge fixed action that one gets is exactly (\ref{e35}).
In other words, for $\alpha \neq \pm{\pi\over 4}$, the action (\ref{e35})
 is actually the dual action  (\ref{e37}) . One can check this by
explicit calculation. Indeed
\begin{eqnarray}
&& S_{LR} - \frac{1}{2\pi} \int d^2 z \; \frac{2 \sin 2 \alpha}{1+B(x)\sin 2
 \alpha} J_L \bar{J}_R \nonumber\\
&& = \frac{1}{1 + B(x) \sin 2 \alpha} \left( ( 1-B(x)\sin 2 \alpha) (\partial
 \theta_L \bar{\partial} \theta_L + \partial \theta_R \bar{\partial} \theta_R )
 - 2 \sin 2 \alpha \partial \theta_L \bar{\partial} \theta_R \right. \\
&& + \left. 2 B(x) \partial \theta_R \bar{\partial} \theta_L + G_{La}(x)
 \partial x^a (\bar{\partial} \theta_L - \sin 2 \alpha \bar{\partial} \theta_R)
 + G_{Ra}(x) \bar{\partial} x^a (\partial \theta_R - \sin 2 \alpha \partial
 \theta_L) \right) \nonumber\\
&& + S^\prime[x]. \nonumber
\end{eqnarray}
By introducing the variables $\theta_0$ and $\theta_1$ through the relations
\begin{eqnarray}
&& \theta_R = \frac{1}{\cos 2 \alpha} ( \sin \alpha \theta_0 + \cos \alpha
 \theta_1 ) \\
&& \theta_L = \frac{1}{\cos 2 \alpha} ( \cos \alpha \theta_0 + \sin \alpha
 \theta_1 ), \nonumber
\end{eqnarray}
which are non-singular for $\alpha \neq \pm \pi /4$, we get exactly
(\ref{e37}).

 The representation (\ref{e35}) of the dual action is
interesting first of all because it  makes the relation with
 marginal perturbations explicit. The integrability of marginal perturbations
is
 also evident if one notes that from (\ref{e37}) the chiral currents of the
 dualized model are given by
\begin{eqnarray}
\bar{J}_R(\alpha) & = & \frac{\cos 2 \alpha}{1 + B(x) \sin 2 \alpha} \bar{J}_R
 \label{elephant} \\
J_L(\alpha) & = & \frac{\cos 2 \alpha}{1 + B(x) \sin 2 \alpha} J_L \nonumber.
\end{eqnarray}
Then, combining (\ref{e35}) and (\ref{elephant}), one immediately gets that
\begin{equation}
S_{dual}(\alpha + \delta \alpha) = S_{dual}(\alpha) - {1\over {2\pi}}
\int d^2z\; \frac{4 \delta
 \alpha}{\cos 2 \alpha} J_L(\alpha) \bar{J}_R(\alpha).
\end{equation}
which is a marginal perturbation around $\alpha\neq 0$.
More interestingly, perhaps,
the action (\ref{e35}) seems to be a natural definition of duality as well as
of
 gauging by a chiral current, in the case $d_L=d_R=1$.
 For $\alpha \neq \pm \pi/4$, (\ref{e35}) obtained from (\ref{e300})
by integrating over the gauge fields,  is the dual
 Lagrangian. At $\alpha = \pm \pi/4$   (\ref{e300}) develops a gauge invariance
  which permits gauging away one of the field, and it turns into the gauged
model $S_{V/A}$. This observation
  puts duality and gauging on the same footing; gauging is duality
 at a point where a gauge invariance develops. This interpretation is probably
 generalizable to the general case of $O(d,d)$ and  to any gauging of abelian
 isometries.

\end{document}